\documentclass{emulateapj}

\begin{document}

\shortauthors{Melso et al.}
\shorttitle{Search for Companions to the Nearest Brown Dwarfs}
\title{A Search for Substellar Companions to the Two Nearest Brown Dwarf
Systems\altaffilmark{1}}

\author{
N. D. Melso\altaffilmark{2}, K. M. Kaldon\altaffilmark{2},
and K. L. Luhman\altaffilmark{2,3}}

\altaffiltext{1}
{Based on data from the {\it Spitzer Space Telescope}
and the ESO Telescopes at Paranal Observatory under program ID 290.C-5195.}

\altaffiltext{2}{Department of Astronomy and Astrophysics,
The Pennsylvania State University, University Park, PA 16802, USA;
kluhman@astro.psu.edu} 

\altaffiltext{3}{Center for Exoplanets and Habitable Worlds, The 
Pennsylvania State University, University Park, PA 16802, USA}

\begin{abstract}

WISE J104915.57$-$531906.1~A+B and WISE J085510.83$-$071442.5
were recently discovered as the third and fourth closest known
systems to the Sun, respectively (2.0 and 2.3~pc).
The former consists of a L8+T0.5 binary 
and the latter is a probable Y dwarf and is the coldest known brown dwarf
($\sim250$~K).
We present a search for common proper motion companions to these brown dwarfs
using multi-epoch mid-infrared images from the {\it Spitzer Space Telescope}.
We have also obtained near-infrared adaptive optics images of
WISE J104915.57$-$531906.1~A+B with the Very Large Telescope to search for
companions at smaller separations than reached by {\it Spitzer}.
No new companions are detected in either system.
At projected separations of 25--420$\arcsec$ (50--840~AU) for
WISE J104915.57$-$531906.1~A+B and 4--420$\arcsec$ (9--970~AU) for
WISE J085510.83$-$071442.5, the {\it Spitzer} images are sensitive to
companions with $M_{4.5}\lesssim21.6$ and 21.9, respectively,
which correspond to masses of $\gtrsim1$~$M_{\rm Jup}$ for ages of $\geq1$~Gyr
and temperatures of $\gtrsim$150~K. The detection limit in the adaptive optics
images of WISE J104915.57$-$531906.1~A+B is $\Delta H\sim10$ at
3--$15\arcsec$ (6--30~AU), or $\gtrsim7$~$M_{\rm Jup}$ for $\geq1$~Gyr.

\end{abstract}


\section{Introduction}
\label{sec:intro}

The all-sky mid-infrared (IR) images obtained by the {\it Wide-field Infrared
Survey Explorer} \citep[{\it WISE},][]{wri10} have enabled the discovery
of a large number of brown dwarfs in the solar neighborhood, particularly
at very low temperatures \citep{cus11,kir11}.
The closest of these newly found brown dwarfs are WISE J104915.57$-$531906.1 A
and B \citep[hereafter WISE 1049$-$5319 A and B,][]{luh13} and
WISE J085510.83$-$071442.5 \citep[hereafter WISE 0855$-$0714,][]{luh14a,luh14b},
which are the third and fourth closest known systems to the Sun, respectively
(2.0 and 2.3~pc).
WISE 1049$-$5319 A and B have spectral types of L8 and T0.5 \citep{luh13,bur13}
and WISE 0855$-$0714 likely has a spectral type of Y \citep{luh14b},
making them the closest known members of their respective spectral classes.
Due to their proximity, these systems are ideal targets for a direct imaging
search for substellar companions at very low luminosities and temperatures.
To search for companions to these brown dwarfs at wide separations
($>5\arcsec$, $>10$~AU), the {\it Spitzer Space Telescope} \citep{wer04} is the
best available telescope because it offers the greatest sensitivity
in the mid-IR bands where cold substellar objects are brightest. For instance,
{\it Spitzer} is capable of detecting a 1~$M_{\rm Jup}$ object with an age
of 1~Gyr at the distances of WISE 1049$-$5319 and WISE 0855$-$0714
\citep{bur03}. Meanwhile, near-IR adaptive optics (AO) images
can be used to search for companions to WISE 1049$-$5319 A and B down to
$0\farcs1$, or 0.2 AU.  WISE 0855$-$0714 has been observed with AO
in the $H$ band by \citet{wri14}, but it was not detected. 
In this paper, we present multi-epoch imaging from {\it Spitzer}
for WISE 1049$-$5319 and WISE 0855$-$0714 and AO imaging from VLT for
WISE 1049$-$5319.

\section{Observations}

\subsection{Near-IR AO Images from VLT} 
\label{sec:ao}

\begin{figure}
\epsscale{0.6}
\plotone{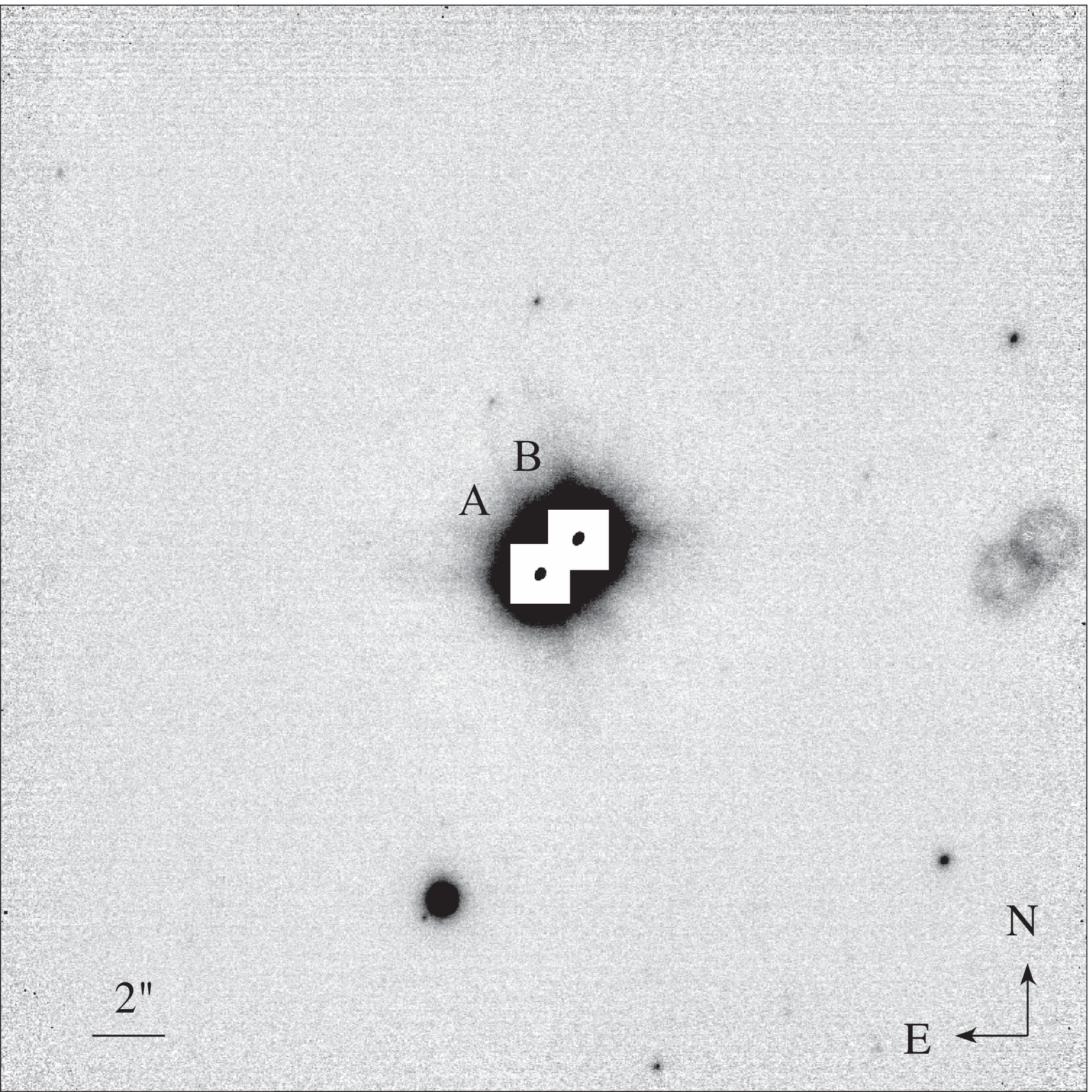}
\caption{VLT NACO $H$-band image of WISE 1049$-$5319~A and B.
To better show the positions of the binary components, we have reduced the 
counts near them by a factor of 30. The size of the image is
$30\arcsec\times30\arcsec$.}
\label{fig:ao}
\end{figure}

Near-IR AO images were used to search for companions to
WISE 1049$-$5319 A and B at small separations. These observations were
performed on the Unit Telescope 4 of the Very Large Telescope (VLT) with the
Nasmyth Adaptive Optics System (NAOS) and the High-Resolution Near-IR Camera
(CONICA), which together are known as NACO \citep{rou03,len03}.
NACO was operated with the S27 camera, the N90C10 dichroic, and the $H$ filter.
The S27 camera contains a 1024$\times$1024 array and has a plate scale of
$0\farcs027$~pixel$^{-1}$, corresponding to a field of view of 
$28\arcsec\times28\arcsec$.
We selected individual exposure times of 4 and 120~sec. The former
provided unsaturated data for the binary components
that could reveal companions at small separations, and the latter provided
greater sensitivity to companions at large separations.
We obtained 10 dithered short exposures and 17 dithered long exposures
on the night of 2013 April 13. In these data, the point spread functions (PSFs)
of the binary components exhibited slight elongations in the direction of the
axis connecting the pair, which was likely caused by the fact
that both objects were present in the wavefront sensor sub-pupils.
Because of these elongations, the observatory repeated the observations on
the night of 2013 May 13, which produced similar results as on the first night.
After performing dark subtraction and flat fielding, we registered
and combined the images at a given exposure time from a given night.
The final combined images from each of the two nights exhibit similar
sensitivity and FWHM ($\sim0\farcs1$). The combined image from the second
night for the 120~sec exposures is shown in Figure~\ref{fig:ao}.
In the long exposures, saturation occurs within the cores of the PSFs
of the binary components ($<0\farcs1$).

\subsection{Mid-IR Images from {\it Spitzer}}
\label{sec:irac}

To search for co-moving companions in wide orbits, we obtained multi-epoch
images of fields surrounding WISE 1049$-$5319 and WISE 0855$-$0714 with
the Infrared Array Camera \citep[IRAC;][]{faz04} on board the
{\it Spitzer Space Telescope}. IRAC has a plate scale of $1\farcs2$ and a
field of view of $5\farcm2\times5\farcm2$. Two filters were available with
IRAC, which were centered at 3.6 and 4.5~\micron\ (denoted as [3.6] and [4.5]).
Because the latter provides better sensitivity to cold brown dwarfs, only
the maps in that band were centered on the targets. We did collect images
at 3.6~\micron\ in flanking fields during the 4.5~\micron\ observations.
WISE 1049$-$5319 was observed on 2013 May 3 and 2013 September 29
through Astronomical Observation Requests (AORs) 48641024 and 48640512,
respectively. WISE 0855$-$0714 was observed on 2014 July 1 and 2015 January 29
through AORs 51040000 and 51040256, respectively.
For each epoch and band for WISE 1049$-$5319, we obtained one short exposure
and one long exposure at each of three dither positions near each of 18
locations in a $6\times3$ grid of pointings separated by 150 and 260$\arcsec$,
respectively. For WISE 0855$-$0714, nine dithered long exposures were
collected near each of nine
positions in a $3\times3$ grid of pointings separated by 260$\arcsec$ in each
direction. For both targets, the long exposure times were 23.6 and 26.8~sec at
3.6 and 4.5~\micron, respectively. A short exposure time of 0.8~sec was used
for WISE 1049$-$5319. The short exposures were included to provide images
in which WISE 1049$-$5319~A and B were not saturated.
These data were reduced in the manner described by \citet{luh12}. 
A combination of the reduced long exposures in both bands and epochs
is shown in Figures~\ref{fig:im1049} and \ref{fig:im0855} for
WISE 1049$-$5319 and WISE 0855$-$0714, respectively.
For each system, a field within 420$\arcsec$ was fully covered
by both epochs at 4.5~\micron, corresponding to 840 and 970~AU,
respectively, given their distances \citep{bof14,luh14pi}.
The components of WISE 1049$-$5319 had a separation of $1\farcs5$ in 2013
\citep{luh13,bur13} and are only partially resolved in these data.

\section{Analysis}
\label{sec:analysis}

\begin{figure}
\epsscale{1.1}
\plotone{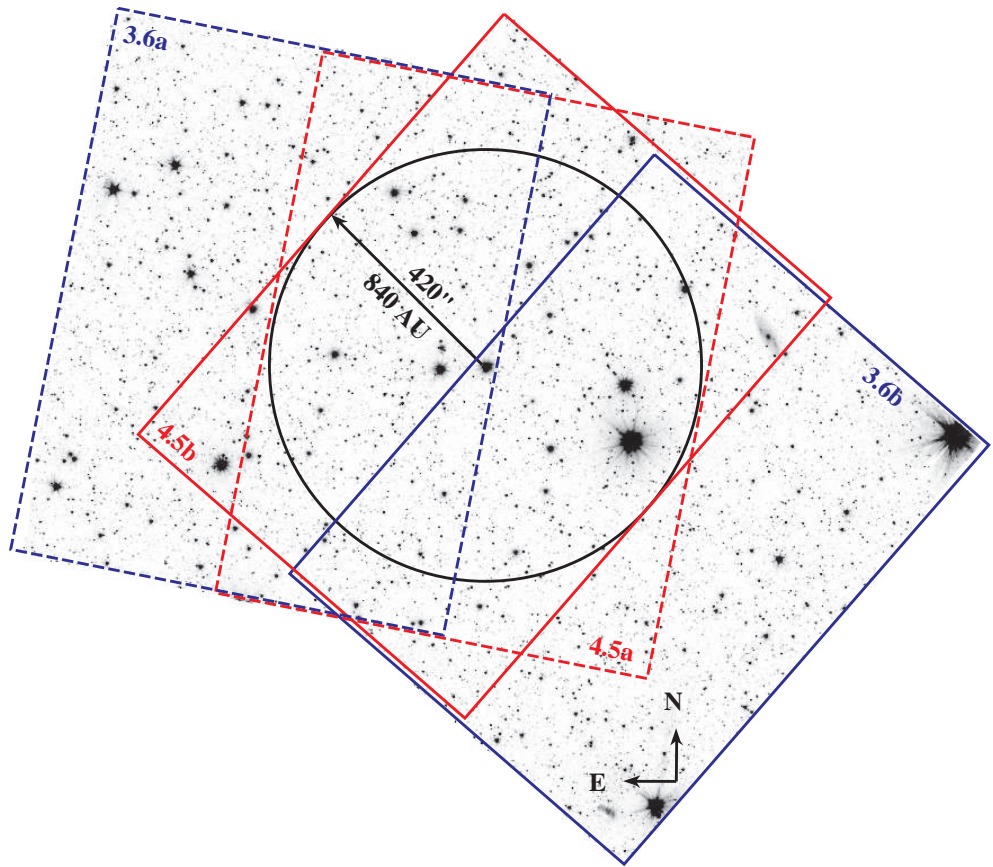}
\caption{Combination of IRAC images of WISE 1049$-$5319 from two bands 
(3.6 and 4.5~\micron) and two epochs (``a" and ``b").
We have searched for common proper motion companions to WISE 1049$-$5319 in
the areas imaged at two epochs. 
The greatest sensitivity to substellar companions is achieved in the
overlapping area between the two epochs at 4.5~\micron,
which provides full coverage out to 420$\arcsec$ (840~AU) from WISE 1049$-$5319
(circle).}
\label{fig:im1049}
\end{figure}

\begin{figure}
\epsscale{0.9}
\plotone{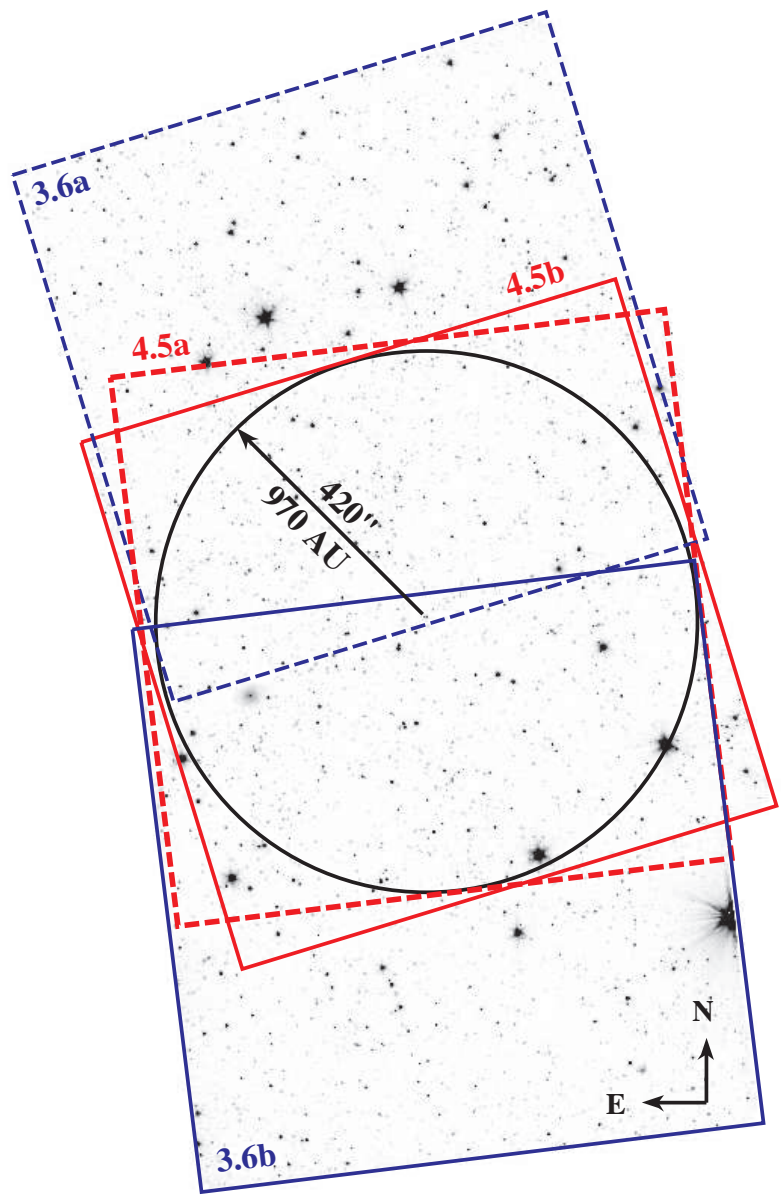}
\caption{Same as Figure~\ref{fig:im1049} for WISE 0855$-$0714.}
\label{fig:im0855}
\end{figure}

\begin{figure}
\epsscale{1.1}
\plotone{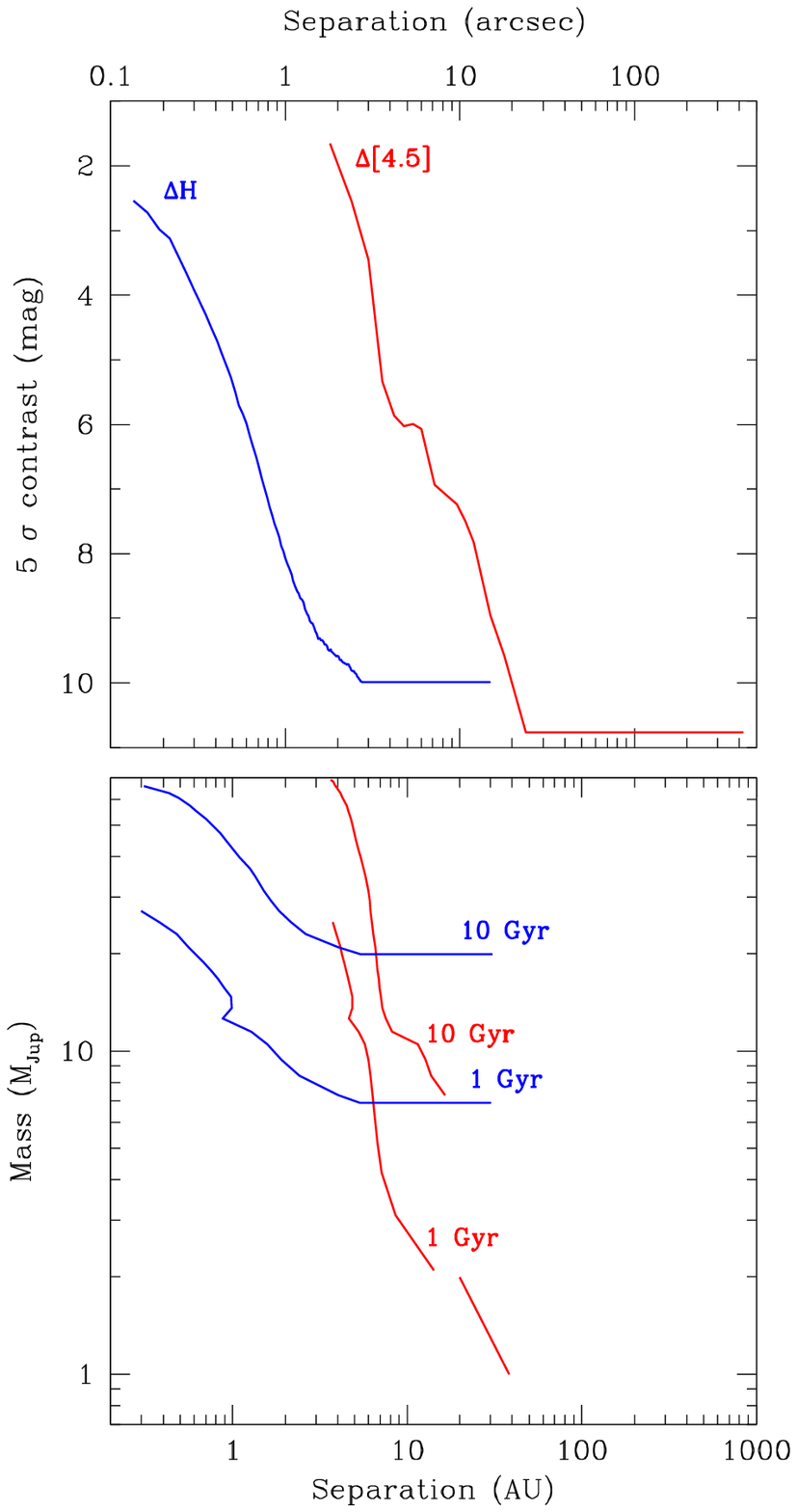}
\caption{
Top: relative magnitude limits (5~$\sigma$) as a function of angular separation
in the AO $H$-band images and IRAC 4.5~\micron\ images of WISE 1049$-$5319~A
and B. Bottom: mass limits as a function of physical separation based on the
measured magnitude limits in the top diagram and the fluxes predicted by
\citet[][$>2$~$M_{\rm Jup}$, 1 and 10~Gyr]{sau12} and
\citet[][1--2~$M_{\rm Jup}$, 1~Gyr]{bur03}.
These images were capable of detecting companions to WISE 1049$-$5319~A and B
down to planetary masses, but no such objects were found.
}
\label{fig:limits}
\end{figure}

Because WISE 1049$-$5319~A and B have similar colors and magnitudes
and appear near the same position in NACO's field of view, we can
use the PSF of one component for PSF subtraction of
the other. The PSF-subtracted versions of the short and long exposures
do not show any additional components at close separations.
Outside of the PSFs of the components, several objects are detected,
as shown in Figures~\ref{fig:im1049} and \ref{fig:im0855}.
None of these sources exhibit a motion between the two epochs that is consistent
with the motion of the binary. Most of these stars are also detected in
$i$-band images from \citet{luh13}, and a comparison of those images with
the NACO data further indicates that they are not co-moving companions.
WISE 1049$-$5319 moved $\sim1\arcsec$ between the $i$ observations and
the second epoch with NACO, but all of the sources detected in both images
remained stationary to within $\sim0\farcs1$. 
To estimate the detection limit for companions in the NACO data, we measured the
standard deviations within annuli across a range of radii from each component.
The width of each annulus was four pixels, which is similar to the FWHM
of the PSF. Because the PSFs of the components overlap, we ignored the data
in the half of each annulus in the direction of the other component.
In other words, the standard deviations were computed for the portions of
the annuli from position angles of 45--225$\arcdeg$ for A and 0--45 and
225-360$\arcdeg$ for B. The standard deviations as a function of separation
are similar for the two stars, which is expected since they have similar
$H$-band magnitudes. We have computed the average of the two curves of
standard deviation versus separation. In the top panel of
Figure~\ref{fig:limits}, we show that average curve in terms
of the 5~$\sigma$ magnitude contrast relative to the
unresolved $H$-band magnitude for the binary system from the Point Source 
Catalog of the Two Micron All-Sky Survey \citep{skr06}.

\begin{figure}
\epsscale{1.1}
\plotone{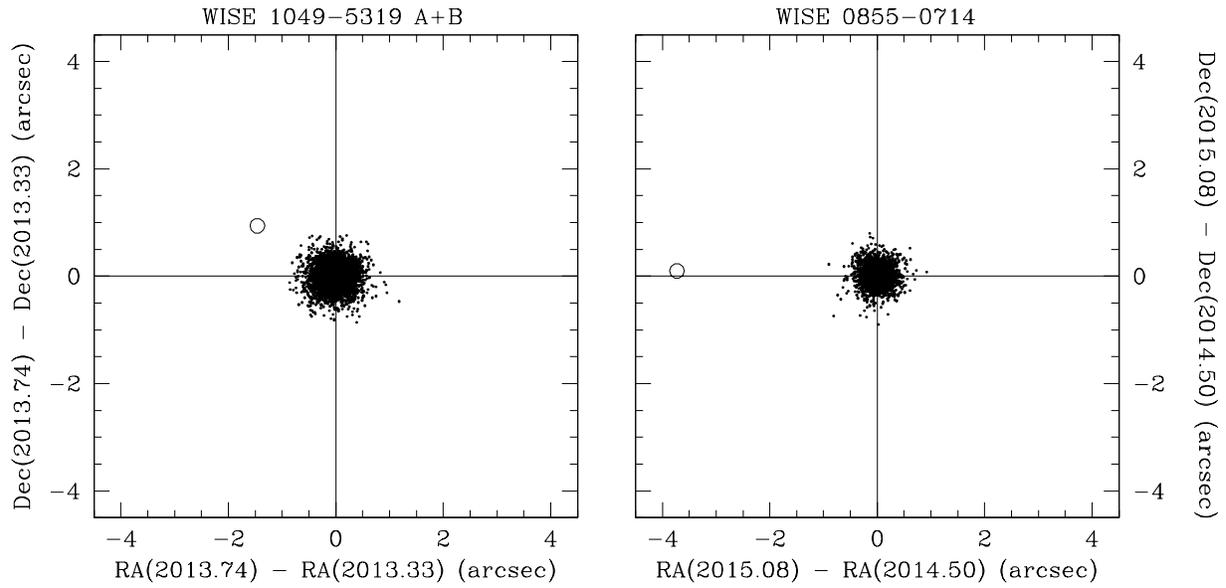}
\caption{Differences in coordinates between two epochs of IRAC images for
WISE 1049$-$5319 A+B and WISE 0855$-$0714 (circles) and all other sources
(points). These images have not detected any objects that share the same
motions as WISE 1049$-$5319 A+B and WISE 0855$-$0714.}
\label{fig:pm}
\end{figure}

To search the IRAC images of WISE 1049$-$5319 and WISE 0855$-$0714
for companions, we began by measuring the positions for all point sources
in each band and epoch with the task {\it starfind} within IRAF.
The resulting positions were transformed to equatorial coordinates using the
World Coordinate Systems in the image headers. We identified the closest
matches between the two epochs for each combination of bands, namely
3.6a/3.6b, 3.6a/4.5b, 4.5a/3.6b, and 4.5a/4.5b where ``a" and ``b" refer
to the two epochs (see Figs.~\ref{fig:im1049} and \ref{fig:im0855}).
The differences in
coordinates for these matches are shown in Figure~\ref{fig:pm}. The motions
of WISE 1049$-$5319 and WISE 0855$-$0714 are large enough that the same
motions for co-moving companions should be easily detected for the faintest
sources in the images, but no such objects are present in Figure~\ref{fig:pm}.
As with the AO data, we estimated the detection limit at 4.5~\micron\ as a 
function of separation from WISE 1049$-$5319~A and B based on the standard
deviations within annuli over a range of radii, where the annuli were given
widths of $1\farcs8$. The resulting values of 5~$\sigma$ are plotted relative
to the combined 4.5~\micron\ magnitude of WISE 1049$-$5319~A and B in the
top panel of Figure~\ref{fig:limits}. Because WISE 0855$-$0714 is much fainter
than WISE 1049$-$5319, the sky background dominates the PSF down to rather
small separations of $\sim4\arcsec$. As a result, the detection limit does not
vary beyond $4\arcsec$ for WISE 0855$-$0714, and hence it is not plotted as a
function of separation in Figure~\ref{fig:limits}. Within $4\arcsec$
from WISE 0855$-$0714, the detection limit in terms of $\Delta$[4.5]
is similar to that of WISE 1049$-$5319.
At separations that are sufficiently large for the sky to dominate,
5~$\sigma$ occurs at $[4.5]=18.1$ and 18.7 for WISE 1049$-$5319 and
WISE 0855$-$0714, respectively, which correspond to $M_{4.5}=21.6$ and 21.9.

We can use evolutionary and atmospheric models of brown dwarfs to convert
the detection limits in $H$ and [4.5] to limits in mass.
Because the ages of WISE 1049$-$5319 and WISE 0855$-$0714 are unknown,
we perform this conversion with the fluxes predicted for ages of
1 and 10~Gyr, which encompass the ages of most stars in the solar neighborhood.
We rely primarily on the fluxes from the models by \citet{sau08} and
\citet{sau12} that are cloudless and employ equilibrium chemistry. Other
models that include clouds and non-equilibrium chemistry produce roughly
similar fluxes in $H$ and [4.5] ($\Delta m\lesssim0.2$)
for the ranges of absolute magnitudes probed by our images
\citep{sau12,mor12,mor14}, and hence the derived mass limits do
not depend significantly on the choice of models. The coldest brown dwarfs
modeled by \citet{sau08} and \citet{sau12} ($\sim200$~K) have $M_{4.5}\sim18$,
whereas the IRAC images approach $M_{4.5}\sim22$ at the distances of our
targets. To transform our limits at $M_{4.5}>18$ to masses, we have adopted the
absolute magnitudes from the models by \citet{bur03} for 1~Gyr, which
extend down to $M_{4.5}=20.65$ (for 1~$M_{\rm Jup}$). Those authors did not
perform calculations for the other age of 10~Gyr that we consider.
After combining the fluxes from the above sets of models
with our measured limits in $H$ and [4.5] for WISE 1049$-$5319, we arrive at
the mass limits that are shown in the bottom panel of Figure~\ref{fig:limits}.
The NACO image provides greater sensitivity at smaller separations
(e.g., 25 and 65~$M_{\rm Jup}$ at 0.4~AU for 1 and 10~Gyr, respectively).
The mass limits for the NACO and IRAC images intersect at $\sim3\farcs5$
($\sim$7~AU). At that separation, both images have limits near 7 and
20~$M_{\rm Jup}$ for 1 and 10~Gyr, respectively. 
Because none of the brown dwarf models that we have considered
are as faint as the $M_{4.5}$ limits reached at large separations, we
are not able to estimate precise values for the lowest masses 
that are detectable in the IRAC images. However, an extrapolation of the
mass limits in Figure~\ref{fig:limits} suggests that the IRAC images are able
to detect companions at large separations from WISE 1049$-$5319 (and
WISE 0855$-$0714) that are slightly below 1~$M_{\rm Jup}$ for 1~Gyr
and $\sim4$~$M_{\rm Jup}$ for 10~Gyr. Such objects would have temperatures of
$\sim$150~K according to the models.

\section{Discussion}
\label{sec:discuss}

Because WISE 1049$-$5319 and WISE 0855$-$0714 are nearby and intrinsically
faint, direct imaging of these systems is sensitive to companions at low
luminosities and small orbital distances.
However, no companions have been detected in our NACO and IRAC data,
which is not surprising given the low binary fractions exhibited by
L and T dwarfs \citep[$\sim$20\%,][references therein]{bur07ppv,abe14}. 
WISE 1049$-$5319 is a binary system (L8+T0.5), and triples composed entirely of 
cool dwarfs are especially rare in direct imaging surveys \citep{bur12,rad13}.
Most L and T dwarf binaries have small separations
\citep[$<20$~AU,][]{bur07ppv}, and the same is true for the small number of
known binaries among late-T and Y dwarfs \citep{gel11,liu11,liu12,dup15}.
As a result, it is unlikely that either WISE 1049$-$5319 or WISE 0855$-$0714
has cool companions beyond the boundaries of our images.
Some brown dwarfs that are discovered in wide-field surveys and initially
appear to be isolated objects are later found to be distant companions to
stars \citep{bur00,sch03,burn09,fah10}, but our two targets do not have
co-moving stellar companions based on the {\it WISE} proper motion surveys
by \citet{luh14a} and \citet{kir14}.
Of course, WISE 1049$-$5319 and WISE 0855$-$0714 may have companions below our
detection limits, especially at small separations.
The components of WISE 1049$-$5319 are sufficiently bright for a search for
close companions through radial velocity and astrometric measurements.
Near-IR imaging with the {\it Hubble Space Telescope} is the only available
option for improving the constraints on the presence of close companions to
WISE 0855$-$0714 given that it is only barely detectable with ground-based
telescopes \citep{fah14}.

\acknowledgements
We acknowledge support from grant NNX12AI47G from the NASA Astrophysics
Data Analysis Program. We thank Caroline Morley and Didier Saumon for
providing their model calculations. 2MASS is a joint project of the University of
Massachusetts and the Infrared Processing and Analysis Center at
Caltech, funded by NASA and the NSF.
The Center for Exoplanets and Habitable Worlds is supported by the
Pennsylvania State University, the Eberly College of Science, and the
Pennsylvania Space Grant Consortium.

\clearpage

\end{document}